\documentclass[12pt,preprint]{aastex}  
\usepackage{psfig}
\usepackage{graphicx}

\advance \voffset by 0.00cm\relax

\begin{document}

\title{Constraints on the binary evolution
 from  chirp mass  measurements  }

 \author{TOMASZ BULIK\altaffilmark{1}, KRZYSZTOF BELCZY\'NSKI\altaffilmark{2,1,3} }

 \affil{$^{1}$ Nicolaus Copernicus Astronomical Center,
       Bartycka 18, 00-716 Warszawa, Poland;\\   
     $^{2}$ Northwestern University, Dept. of Physics \& Astronomy,
       2145 Sheridan Rd., Evanston, IL 60208\\
     $^{3}$ Lindheimer Postdoctoral Fellow\\  
     bulik@camk.edu.pl, belczynski@northwestern.edu}

 \begin{abstract}   
We estimate the observed distribution of chirp masses
of compact object binaries for  the gravitational wave detectors. 
The stellar binary evolution is modeled using
 the {\em StarTrack} population synthesis code.
The distribution of the predicted "observed" chirp masses
vary with variation of different parameters describing
stellar binary evolution. We estimate the sensitivity 
of the observed distribution to variation of these parameters
and show which of the parameters can be constrained after observing
20, 100, and 500 compact object mergers. As a general feature of all our models
we find that the  population of observed binaries is dominated by the
double black hole mergers.
 \end{abstract}

\keywords{binaries: close --- gravitational waves}

\newcommand\chirp{{\cal M}}

\section{INTRODUCTION}

Compact object mergers are one of the most promising sources 
of gravitational waves for the ground based interferometric
detectors like LIGO (Abramovici et al 1992) and VIRGO
(Bradaschia etal 1990). So far most of the theoretical papers
on the properties of these sources related to the 
gravitational wave detections
have been concentrated on calculation of the predicted rates 
(Narayan etal 1991, Phinney 1991,
Kalogera etal. 2001). In this paper we wish to address another aspect
of the gravitational wave detection, i.e. the distribution
of observed masses of the compact objects.

Stellar mass compact object binaries shall be detected 
during the inspiral phase, while the consecutive 
merger and ringdown phases will most likely have lower signal to noise ratios.
During the inspiral phase the motion of the binary components 
and also the wave form is governed by the
chirp mass $\chirp=(m_1+m_2)^{2/5}(m_1 m_2)^{3/5}$ (Peters and Matthews, 1963). 
The waveform will depend on the individual masses of the binary components
$m_1$ and $m_2$ when the post Newtonian effects are taken into account.
However, the analysis of the inspiral phase alone shall not suffice
to determine if a binary contained a neutron star or a black hole without
the prior knowledge of the neutron star maximum mass.
A careful modeling of the signal may yield the individual masses
of the objects, however,  the chirp mass will be the primary observable
for the compact object mergers (Cutler \& Flanagan 1994).

 We use the {\em StarTrack} population synthesis 
code (Belczynski, Kalogera \& Bulik\ 2002) to calculate 
the distributions of compact object binary masses, and we present these  
calculations in \S~2. In \S~3 we estimate the number
of merger detections required to distinguish between 
different models of stellar binary evolution.
Finally,  \S~4 contains the conclusions and discussion.

\begin{table}
\caption{Description of different population synthesis models
for which the distributions of mass ratios have been found.}

\begin{center}
\begin{tabular}{lp{10.7cm}}
\hline \hline
Model & Description \\
\hline
A      & standard model described in  Belczynski, Kalogera, Bulik (2002), 
        but with $T_{Hubble}=15$Gyrs \\
B1,7,11 & zero kicks, single Maxwellian with  
$\sigma=50,500,$\,km\,s$^{-1}$, \\
B13 &   Paczynski (1990) kick with $V_k=600$km\,s$^{-1}$\\
C      & no hyper--critical accretion onto NS/BH in CEs \\
E1--3  & $\alpha_{\rm CE}\times\lambda = 0.1, 0.5, 2$ \\
F1--2  & mass fraction accreted: f$_{\rm a}=0.1, 1$ \\
G1--2  & wind changed by\ $f_{\rm wind}=0.5, 2$ \\
J      & primary mass: $\propto M_1^{-2.35}$ \\
L1--2  & angular momentum of material lost in MT: $j=0.5, 2.0$\\
M1--2  & initial mass ratio distribution: $\Phi(q) \propto q^{-2.7}, q^{3}$\\
O      & partial fall back for $5.0 < M_{\rm CO} < 14.0 \,M_\odot$\\
S      & all systems formed in circular orbits\\
Z1--2  & metallicity: $Z=0.01$, and $Z=0.0001$\\

\end{tabular}
 
\end{center}
\end{table}

\section{Distribution of the chirp masses}

\begin{figure*}[t]

\plottwo{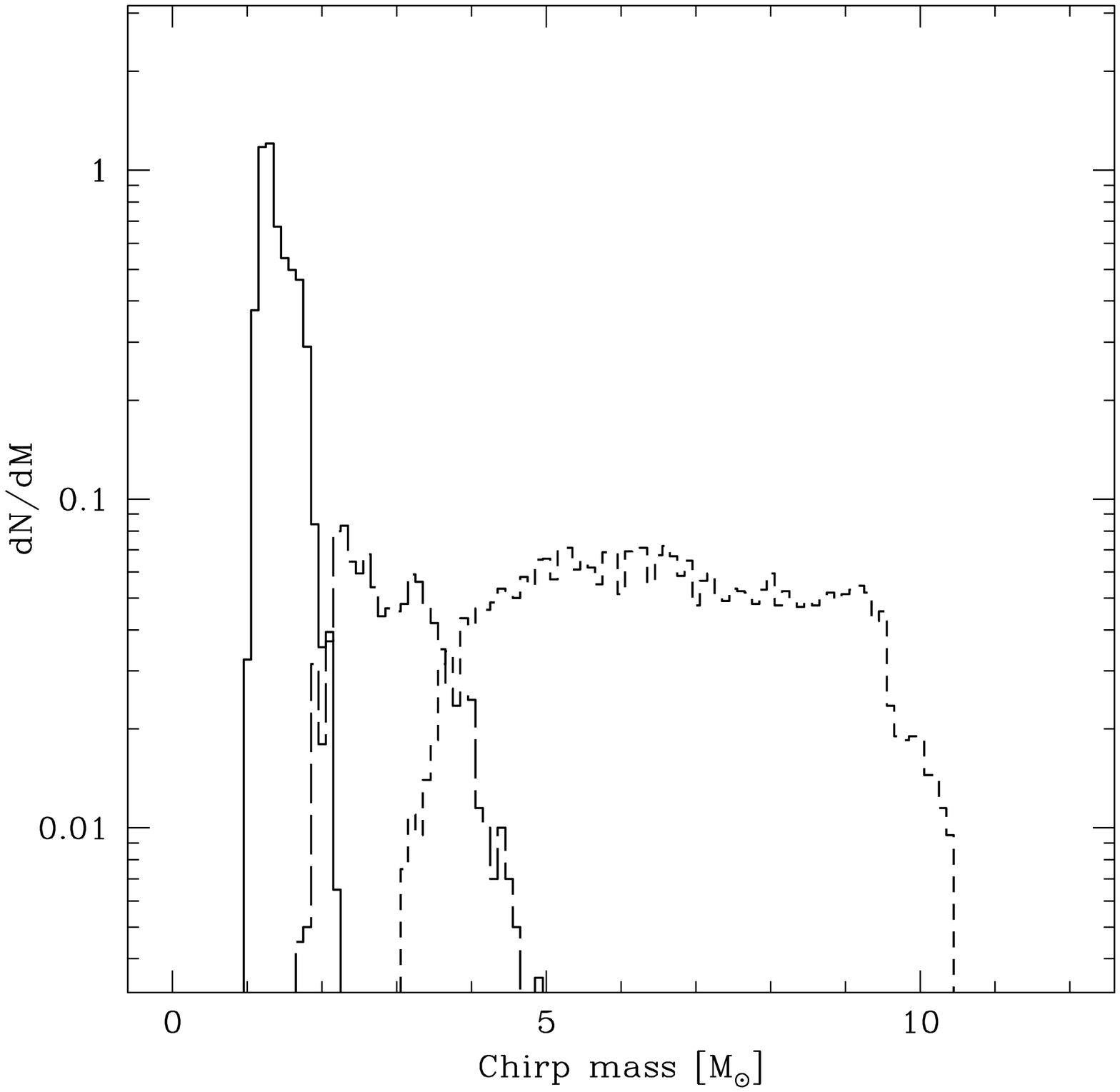}{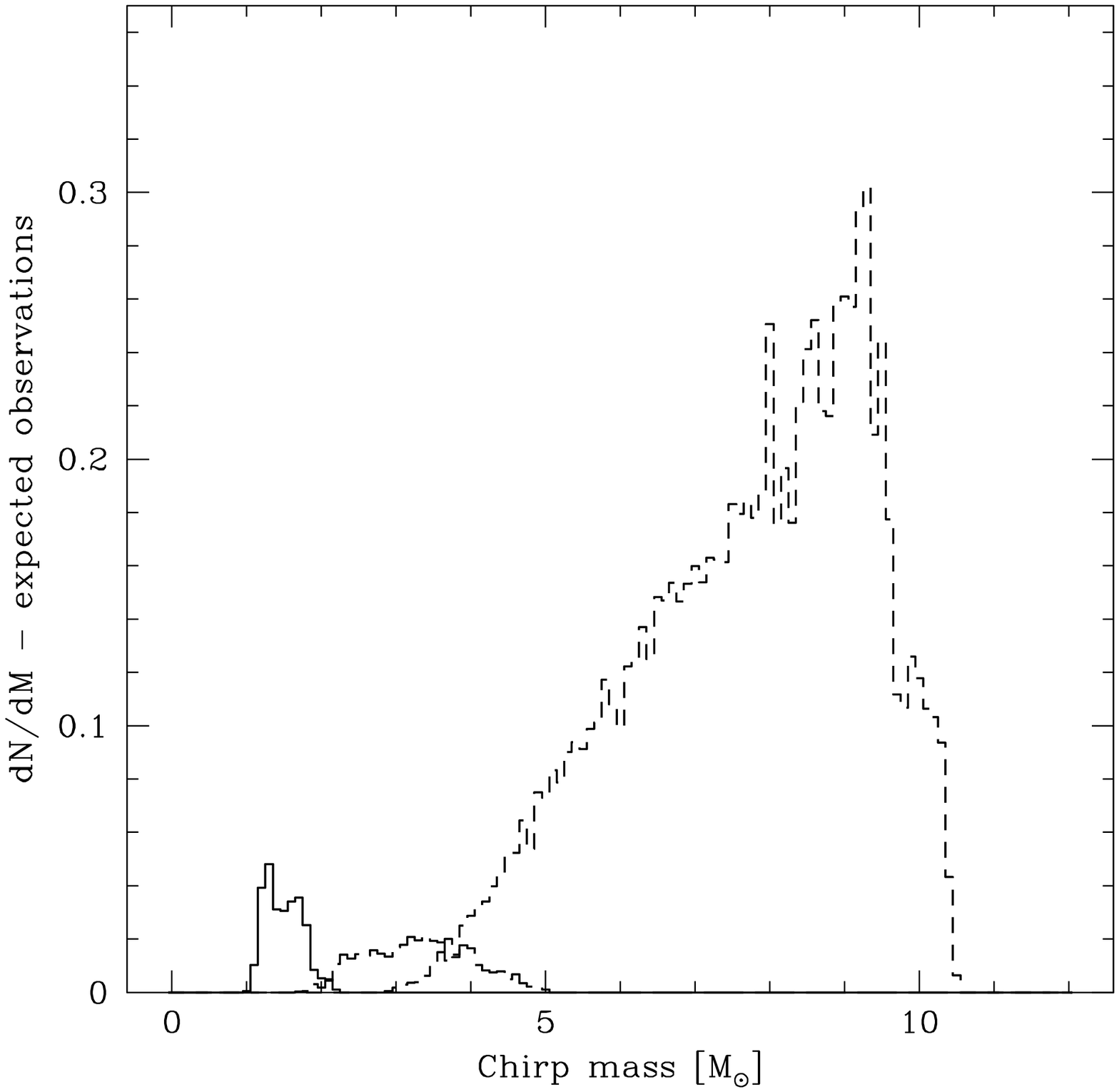}
\caption
{The intrinsic (galactic) distribution of the chirp masses
in the framework of model A (left panel), and 
the  distribution of the expected observations (right panel).
The solid line corresponds to the NS-NS mergers, the short dashed 
line represents the
NS-BH mergers and the dashed line stands for the BH-BH margers.
The sum of the three distribution in each panel is normalized to unity.
}
\label{chirpgal}
\end{figure*}

The StarTrack binary population synthesis code is described
in detail in  Belczynski, Kalogera \& Bulik\ (2002). 
 One of the
important features of the code is the possibility to conduct  parameter
study of a given property of the population of binaries, i.e. 
to estimate the dependence of the result on each of the parameters used 
to describe the stellar and binary evolution. The models
used in this paper are listed in Table 1. We first use 
the standard model A results to present the intrinsic
distribution of the chirp masses. This is shown in Figure~\ref{chirpgal}.
The distribution shows a clear peak at low chirp masses 
$1 M_\odot\chirp < 2 M_\odot$ which 
is due to the double neutron star systems. The mixed (BH-NS) systems 
populate the intermediate region, while the chirp masses of the  BH-BH 
binaries  extend up to above $10 M_\odot$.

In order to estimate the observed distribution of the chirp masses
of compact objects one has to take into account the sensitivity 
of the gravitational wave detectors to signals from mergers
of different binaries. The calculations of the signal to noise
ratio (Finn and Chernoff 1993, Bonazzola and Marck 1994, 
Flanagan and  Hughes 1997) show that 
the sampling distance in the first approximation 
is a function  of the chirp mass only: $D\propto \chirp^{5/6}$. The additional
corrections due to limited sensitivity window of the detectors have been
calculated by Flanagan and Hughes (1997) and amount to approximately 10\% 
for the binaries with the total mass below $18 M_\odot$ for the initial 
LIGO, and less for the 
advanced LIGO. In this paper we neglect these 
corrections. The distribution of the expected observed chirp masses
can be calculated using Monte Carlo method. We assume that the 
Universe is uniformly filled with merging binaries, and 
for each merger we estimate the signal to noise ratio in the detector.
We model the   population of merging binaries  assuming a continuous 
star formation rate. 
The result is shown in the right  panel of Figure~\ref{chirpgal}.
One can note that these distribution could also be obtained 
analytically by multiplying
the distributions of Figure~\ref{chirpgal} by the volume $\propto \chirp^{5/2}$
and normalizing it.
In this plot the BH-BH systems are now the dominant contribution
of the distribution. This is due to the fact that the sampling volume for the 
BH-BH binaries more than  100 times larger that that for the NS-NS
systems, which easily compensates for the lower merger rate of 
the BH-BH binaries.

\section{Expected observations}

\begin{figure*}[t]

\plotone{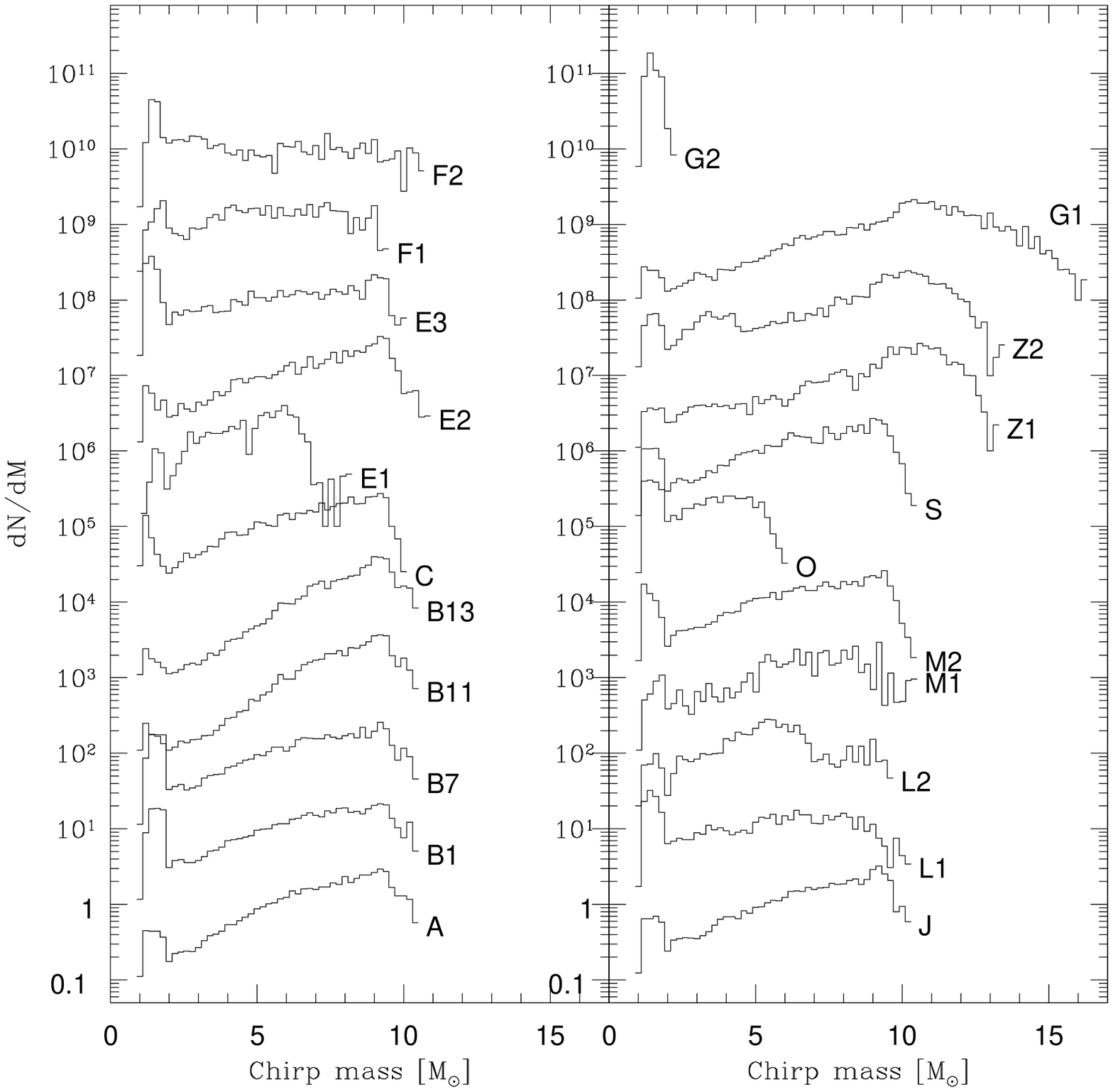}
\caption
{Distributions of the expected observed  chirp masses
in the framework of  models listed in Table~1. For clarity each distribution 
is shifted up by a factor of ten.
}
\label{chidist}
\end{figure*}

Let us now address the following questions: 
are the distributions of observed chirp masses 
expected in the framework of alternative models 
different? If so, are these differences significant?
We simulate the distributions of chirp masses
in the expected observations with the binary populations obtained 
from the set of models of Table~1 similarily
as we have done above for the model A.
We present the results in Figure~\ref{chidist}.
Different stellar evolution models lead to drastically different distributions
of the  chirp masses in the expected observations.
Various parameters describing stellar evolution affect 
the distribution of observed chirp masses in several  ways.
Changing the kick velocity distribution (models B) alters 
the ratio between the number of the neutron star binaries and the back hole
binaries. Other models change the maximal masses
of the black holes produced. This is especially clear in the case
of models G where the stellar winds are varied by a factor of two
upwards (model G2) and downwards (model G1)
We note that  the shapes of these distributions do not depend
on the sensitivity of a detector.

\begin{figure*}[t]

\plotone{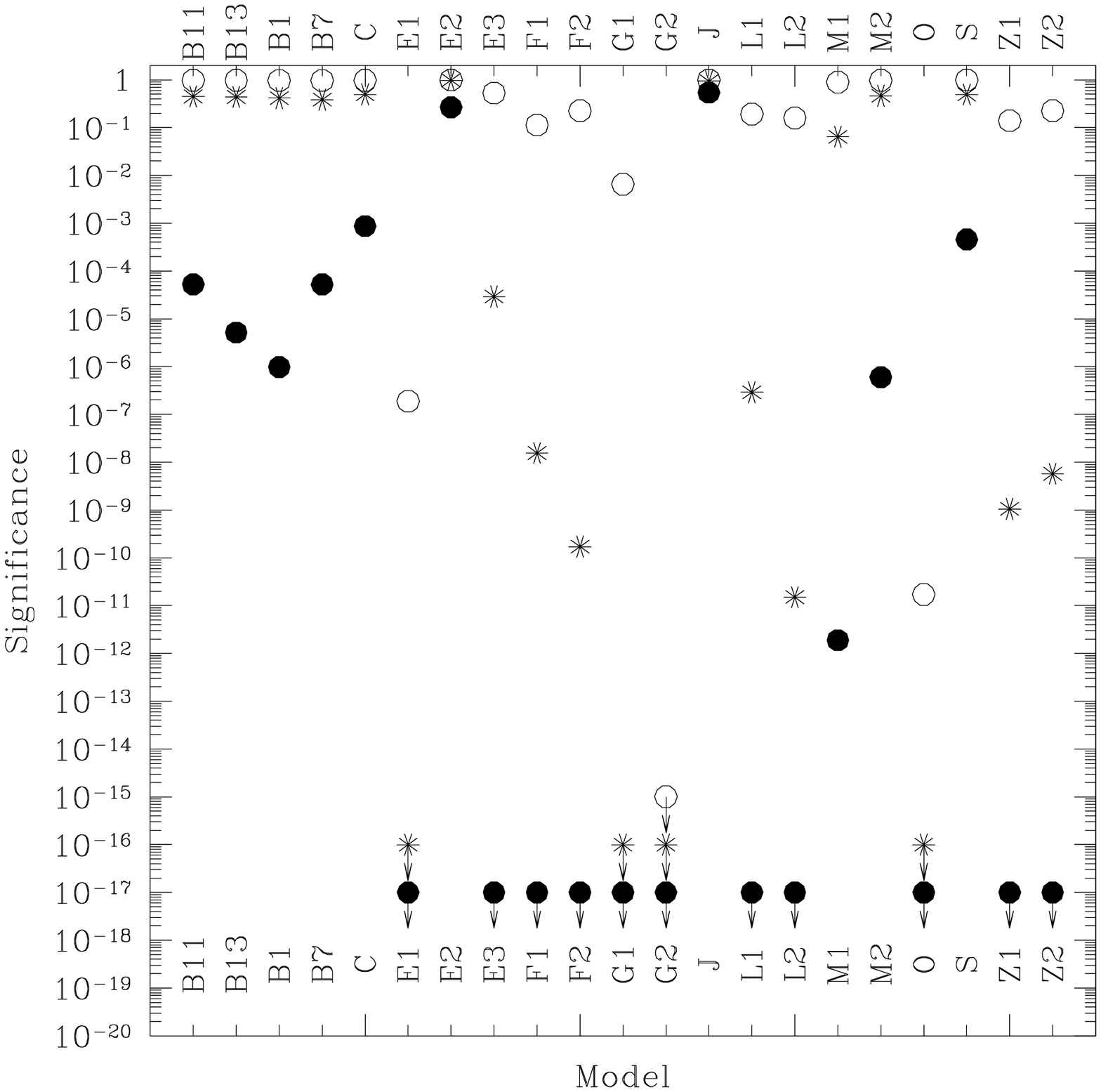}
\caption
{The significance of rejecting model A. Open circles correspond to 
observations of just 20 mergers, stars to 100 mergers, and 
filled circles to observations of 500 mergers. The symbols with 
an arrow denote the case when significance is off the scale.
}
\label{res}
\end{figure*}

In order to verify if the differences between the distributions
are significant we turn to a  simulation of a finite 
number of merger observations. We assume that 
the true stellar evolution goes through one of the models 
of Table~1. We then simulate the observations of a given 
number of mergers (we use 20, 100 and 500 mergers)
and for each such simulated observation
we verify  using the Kolmogorov Smirnov (KS) test if we can 
reject a hypothesis that the stellar evolution is described by 
model A. This allows to test the sensitivity of the 
shape of the distribution of expected chirp mass observations
to the underlying model parameters describing stellar evolution.
For each number of merger observations we
repeat this test 10000 times to obtain a distribution of 
KS-test probabilities and find the lowest probability that appeared 
in the top one percentile of this distribution.
We can now set a detection confidence level, say at $10^{-5}$
and compare each probability with this value:
if it is higher  we conclude that 
this particular model cannot be distinguished from 
model A with  a given number of merger observations, while
a smaller number means that this model can be distinguished, and
that some constriants can be iposed on the particular
parameter through which this model differes from model A.
We present the results of the test in Figure~\ref{res}.

Figure~\ref{res} presents  a measure of sensitivity 
of the expected observed distribution
of chirp masses to the parameters describing stellar evolution. 
One can see from Figure~\ref{res} that even observations of a small 
number of mergers (open circles correspond to 20 mergers)
yield highly significant results 
for models E1, G2 and O. The reason for that can be is clear from 
Figure~\ref{chidist}. These are the models for which the maximal 
chirp mass in the population is significantly lower than that for model A.
Model G2 population (i.e. with increased stellar winds) contains
hardly any black holes. In general we see that  these observations
are very sensitive to the value of maximum mass of stellar 
black holes in the population. Model G1 (with decreased stellar winds) 
which allows for formation of black holes binaries with chirp masses
up to $16\,M_\odot$will stand out with less than a hundred merger
observations.

With a larger number of  merger observation (stars in Figure~\ref{res})
correspond to 100 merger detections) more parameters
can be constrained. Some constraints can be obtained for 
the value of  common envelope 
efficiency $\alpha_{CE}\lambda$ (models E: model E2 is similar to A).
Other parameters describing mass transfer events like
 the mass fraction accreted (models F), and the amount of angular 
momentum lost (models L) shall also be constrained. Moreover 
also the metallicity of the progenitor stars may influence 
the observed distribution at a significant level (models Z).

Constraining the initial mass ratio distribution (models M)
will require an even higher number of merger detections: only for
the case of 500 observed mergers the differences become significant.
Models C (no hypercritival accretion onto a compact object), 
J (initial mass function slope), and S (systems circular initially)
lead to very small differences in the 
observed distribution of chirp masses.

Models B where the kick velocity distribution is varies begin to show
significant differences only with a large number of observations.
Changing the kick velocity distribution affects strongly the absolute rates
(Lipunov Postnov Prokhorov 1997, Belczynski Bulik  1999),
and the ratio of double neutron star mergers to the double black hole
mergers (Belczynski, Kalogera and Bulik 2002).

\section{CONCLUSIONS}

We have applied the stellar population synthesis models
to simulate the distribution of observed 
chirp masses in the gravitational wave detection of stellar mergers.
We find that the population of observed mergers is dominated 
by the the black hole - black hole binary mergers. In most models 
double black hole mergers constitute more than 90\% 
of the observed events. The exception is model G2, in which the formation of
black holes is suppressed because of increased stellar winds.
The shapes of observed distributions of chirp masses vary considerably
for  different models of stellar binary evolution.

We simulate the observed distributions of chirp masses
in the framework of various stellar evolution models and estimate the
sensitivity with which these parameters can be estimated from a given sample
of observed mergers. We find that there is  large number 
of parameters that can be constrained given a set of measured 
chirp masses. The main and immediate constraints come from the fact
the the observed population seems to be dominated by the highest mass black hole
binaries. Thus even a small set of observations yields constraints on 
the maximal mass of merging black hole binaries. A larger set of observations 
will lead to constraints on the evolution of high mass
binaries.

In our simulation we use a simple statistical tool:
the Kolmogorov Smirnov test. Given a set of real observations with 
some measurements of individual masses of coalescing stars,
one could use a more sensitive tool  like the maximum likelihood
method. However, even with such simple statistic as used here we can show the 
general properties of the expected observations, and demonstrate 
the sensitivity of the observed distributions to different model parameters.

we note that consideration of the distribution of observed masses 
will lead to stricter constraints than consideration of just the 
observed rates. The theoretical calculation of rates 
involves estimating a number selection effects and calibrating 
with other sources which leads to several uncertainties.
 The calculation of the observed 
distribution of chirp masses is free from such uncertainties because
a distributions essentially equivalent to considering the ratios 
of the number of mergers of different type, and all the normaliztion 
 factors that enter the rate estimates do cancel out.

Finally, we have to mention several effects not taken into account
in this paper. A more detailed  calculation must include the consideration
of the effects of lifetimes of binaries of different type. 
Belczynski, Kalogera and Bulik (2002) have shown
that the typical lifetimes of double neutron star binaries are 
much smaller than the black hole binaries. The effects due to 
changing of the star formation rate with the redshift 
will affect the observed population of merging binaries.
When considering the advanced detectors sensitive to mergers 
at cosmological distances one also needs to take into account
cosmological effects: the fact that the true measure quantity is the redshifted
mass $(1+z)\chirp$, and also the change of the observed volume 
with the redshift (here we have assumed the Euclidean geometry). 
These issues will be considered  in a separate paper.

\acknowledgements We  thank Vicky Kalogera 
for comments on this project.  TB thanks the hospitality
of the Theoretical Astrophysics Group at Northwestern University.
We acknowledge support from KBN through grant 5P03D01120.

\end{document}